\documentclass{article}

\usepackage{spconf}
\usepackage{amsmath}
\usepackage{graphicx}

\usepackage{microtype}
\usepackage{graphicx}

\usepackage{subcaption}

\usepackage{booktabs} 
\usepackage{xcolor} 
\usepackage{pgfplots,pgfplotstable}

\usepackage{hyperref}
\usepackage{url}

\usepackage{amsmath}
\usepackage{amssymb}
\usepackage{mathtools}
\usepackage{amsthm}
\usepackage{bm}
\usepackage[capitalize,noabbrev]{cleveref}

\definecolor{forestgreen}{rgb}{0.13, 0.55, 0.13}
\definecolor{fulvous}{rgb}{0.86, 0.52, 0.0}
\definecolor{glaucous}{rgb}{0.38, 0.51, 0.71}
\definecolor{lava}{rgb}{0.81, 0.06, 0.13}
\definecolor{buff}{rgb}{0.94, 0.86, 0.51}
\definecolor{chromeyellow}{rgb}{1.0, 0.65, 0.0}
\definecolor{brightube}{rgb}{0.82, 0.62, 0.91}

\hypersetup{
    pdfborderstyle={/S/U/W 1}, 
    linkbordercolor=red,       
    citebordercolor=green,     
    filebordercolor=magenta,   
    urlbordercolor=cyan        
}

\pgfplotsset{scaled y ticks=false, scaled x ticks=false}

\theoremstyle{plain}

\theoremstyle{definition}

\theoremstyle{remark}

\usepackage{xspace}
\newcommand{\RADMMM}{RM\xspace}
\newcommand{\RADTTSML}{RT\xspace}
\newcommand{\TC}{T2\xspace}

\usepackage[textsize=tiny]{todonotes}

\usepackage{siunitx}

\usepackage{float}
\usepackage{adjustbox}
\title{Multilingual Multiaccented Multispeaker TTS with RADTTS}

\name{Rohan Badlani, Rafael Valle, Kevin J. Shih, João Felipe Santos, Siddharth Gururani, Bryan Catanzaro}

\address{NVIDIA}

\begin{document}
\maketitle
\begin{abstract}

\vspace*{-0.25\baselineskip}
We work to create a multilingual speech synthesis system which can generate speech with the proper accent while retaining the characteristics of an individual voice. This is challenging to do because it is expensive to obtain bilingual training data in multiple languages, and the lack of such data results in strong correlations that entangle speaker, language, and accent, resulting in poor transfer capabilities. To overcome this, we present a multilingual, multiaccented, multispeaker speech synthesis model\footnote{Samples can be found at \href{https://drive.google.com/drive/folders/1PTt1OSum11PASOA7-YmgE87bO6uddcPx?usp=sharing}{\textbf{\underline{this link}}}} based on RADTTS with explicit control over accent, language, speaker and fine-grained $F_0$ and energy features. Our proposed model does not rely on bilingual training data. We demonstrate an ability to control synthesized accent for any speaker in an open-source dataset comprising of 7 accents.  
Human subjective evaluation demonstrates that our model can better retain a speaker's voice and accent quality than controlled baselines while synthesizing fluent speech in all target languages and accents in our dataset.
\vspace*{-0.5\baselineskip}
\end{abstract}
\section{Introduction} \label{sec:introduction}
\vspace*{-0.75\baselineskip}
Recent progress in Text-To-Speech (TTS) has achieved human-like quality in synthesized mel-spectrograms \cite{Tacotron, Tacotron2, Flowtron, radtts} and waveforms\cite{NaturalSpeech, kim2021vits}.
Most models support speaker selection during inference by learning a speaker embedding table\cite{Tacotron, Tacotron2, Flowtron} during training, while some support zero-shot speaker synthesis by generating a speaker conditioning vector from a short audio sample\cite{speakerverificationtransfer}. However, most models support only a single language. This work focuses on factorizing out speaker and accent as controllable attributes, in order to synthesize speech for any desired combination of speaker, language and accent present in the training dataset. 

It is very expensive to obtain bilingual datasets because most speakers are unilingual. Hence, speaker, language, and accent attributes are highly correlated in most TTS datasets. Training models with such entangled data can result in poor language, accent and speaker transferability. Notably, every language has its own alphabet and most TTS systems use different symbol sets for each language, sometimes even separate encoders\cite{Nachmani2019UnsupervisedPT}, severely limiting representational sharing across languages. This aggravates entanglement of speaker, language and text, especially in datasets with very few speakers per language. Approaches like \cite{learningfluently} introduce an adversarial loss to curb this dependence of text representations on speaker. Other approaches use a union of linguistic feature sets of all languages\cite{bo2016sharedrepresentations} to simplify text processing for multi-language training. However, these solutions don't support code-switching situations where words from multiple languages appear in mixed order in the synthesis prompt. 



Recently, there has been an interest in factorizing out fine-grained speech attributes\cite{lancucki2021fastpitch,ren2020fastspeech,radpp} like $F_0$ and energy. We extend this fine-grained control by additionally factorizing out accent and speaker with an ability to predict frame-level $F_0$ and energy for a desired combination of accent, speaker and language. We analyze the effects of such explicit conditioning on fine-grained speech features on the synthesized speech when transferring a voice to other languages.


Our goal is to synthesize speech for a target speaker in any language with a specified accent. Related methods include YourTTS\cite{yourtts}, with a focus on zero-shot multilingual voice conversion. Although promising results are presented for a few language combinations, it shows limited success on transferring from languages with limited speakers. Moreover, it uses a curriculum learning approach to extend the model to new languages, making the training process cumbersome. Closest to our work is \cite{learningfluently}, which describes a multilingual and multispeaker TTS model without requiring individual speakers with multiple language samples. 


In this work, we \textbf{(1)} demonstrate effective scaling of single language TTS to multiple languages using a shared alphabet set and alignment learning framework\cite{radtts,onettsalignment}; \textbf{(2)} introduce explicit accent conditioning to control the synthesized accent; \textbf{(3)} propose and analyze several strategies to disentangle attributes (speaker, accent, language and text) without relying on parallel training data (multilingual speakers); and \textbf{(4)} explore fine-grained control of speech attributes such as $F_0$ and energy and its effects on speaker timbre retention and accent quality. 
 
\vspace*{-1.0\baselineskip}
\section{methodology}\label{sec:methodology}
We build upon RADTTS\cite{radtts,radpp} as deterministic decoders tend to produce oversmoothed mels that require vocoder fine-tuning. Our model synthesizes mels($X \in \mathbb{R}^{C_{\mathit{mel}} \times F}$) using encoded text($\Phi \in \mathbb{R}^{C_{\mathit{txt}} \times T}$), accent($A \in \mathbb{R}^{D_{\mathit{accent}}}$) and speaker($S \in \mathbb{R}^{D_{\mathit{speaker}}}$) as conditioning variables, with optional conditioning on fundamental frequency($F_0 \in \mathbb{R}^{\mathit{1} \times F}$) and energy($\xi \in \mathbb{R}^{\mathit{1} \times F}$), where $F$ is the number of mel frames, $T$ is the text length, and energy is the per-frame mel energy average. We propose the following novel modifications: 
\vspace*{-0.8\baselineskip}
\subsection{Shared text token set}
\vspace*{-0.5\baselineskip}
Our goal is to train a single model with the ability to synthesize a target language with desired accent for any speaker in the dataset. We represent phonemes with the International Phonetic Alphabet (IPA) to enforce a shared textual representation. A shared alphabet across languages reduces the dependence of text on speaker identity, especially in low-resource settings (e.g. 1 speaker per language) and supports code-switching. 
\vspace*{-0.8\baselineskip}
\subsection{Scalable Alignment Learning}
\vspace*{-0.5\baselineskip}
We utilize the alignment learning framework in\cite{radtts,onettsalignment}, to learn speech-text alignments $\Lambda \in\mathbb{R}^{T \times F}$ without external dependencies. A shared alphabet set simplifies this since alignments are learnt on a single token set instead of distinct sets. However, when the speech has a strong accent, the same token can be spoken in different ways from speakers with different accents and hence alignments can become brittle. To curb this multi-modality, we learn alignments between (text, accent) and mel-spectograms using accent $A$ as a conditioning variable.
\vspace*{-0.8\baselineskip}
\subsection{Disentangling Factors}
\vspace*{-0.5\baselineskip}
We focus on non-parallel data with a speaker speaking 1 language which typically has text $\Phi$, accent $A$ and speaker $S$ entangled. We evaluate strategies to disentangle these attributes:

\noindent{\textbf{Speaker-adversarial loss}}
In TTS datasets, speakers typically read different text and have different prosody. Hence, there can be entanglement between speaker $S$, text $\Phi$ and prosody. Following\cite{learningfluently}, we employ domain adversarial training to disentangle $S$ and $\Phi$ by using a gradient reversal layer. We use a speaker classification loss, and backpropagate classifier's negative gradients through the text encoder and token embeddings.

\vspace*{-1.6\baselineskip}
\begin{align}\label{eq:adv}
 L_{adv} = \sum_{i=1}^{N} {P(s_i | \phi_i ; \theta_{spkclassifier})} 
\end{align}

\vspace*{-0.8\baselineskip}
\noindent{\textbf{Data Augmentation}}
Disentangling accent and speaker is challenging, as a speaker typically has a specific way of pronouncing words and phonemes, causing a strong association between speaker and accent. Straightforward approaches to learning from non-parallel data learn entangled representations because a speaker's language and accent can be trivially learned from the dataset. Since our goal is to synthesize speech for a speaker in a target language with desired accent, disentangling speaker $S$ and accent $A$ is essential, otherwise either speaker identity is not preserved in the target language or the generated speech retains the speaker's accent from the source language. To overcome this problem, we use data augmentations like formant, $F_0$, and duration scaling to promote disentanglement between speaker and accent. For a given speech sample $x_i$ with speaker identity $s_i$ and accent $a_i$, we apply a fixed transformation $t \in \{1, 2, ... \tau\}$ to construct a transformed speech sample $x_i^{t}$ and assign speaker identity as $s_i + t \cdot N_{speakers}$ and accent as original accent $a_i$, where $\tau$ is the number of augmentations. This creates speech samples with variations in speaker identity and accent in order to orthogonalize these attributes. 

\noindent{\textbf{Embedding Regularization}}
Ideally, the information captured by the speaker and accent embeddings should be uncorrelated. 
To promote disentanglement between accent and speaker embeddings, we aim to decorrelate the following variables: (1) random variables in accent embeddings; (2) random variables in speaker embeddings; (3) random variables in speaker and accent embeddings from \emph{each other}. While truly decorrelating the information is difficult, we can promote something close by using the constraints from VICReg\cite{VicReg}. We denote $E^{A} \in \mathbb{R}^{D_a\times N_a}$, $E^{S} \in \mathbb{R}^{D_s\times N_s}$ as the accent and speaker embedding tables respectively. Column vector $e^j \in E$ denotes the $j$'th embedding in either table. Let $\mu_E$ and $Cov(E)$ be the means and covariance matrices. By using VICReg, we constrain standard deviations to be at least $\gamma$ and suppress the off-diagonal elements of the covariance matrix ($\gamma=1, \epsilon=1e-4$):
\vspace*{-1.5\baselineskip}
\begin{align}\label{eq:var}
 L_{var} &= \frac{1}{D}\sum_{i=j}\max\left(0, \gamma - \sqrt{Cov(E)_{i,j} + \epsilon}\right)\\
 L_{covar} &= \sum_{i\neq j} {Cov(E)_{i,j}^2}
\end{align}
\vspace*{-1.0\baselineskip}

Next, we attempt to decorrelate accent and speaker variables from \emph{each other} by minimizing the cross-correlation matrix from batch statistics. Let $\tilde{E^A}$ and $\tilde{E}^S$ be the sampled column matrices of accent and speaker embedding vectors sampled within a batch of size $B$. We compute the batch cross -correlation matrix $R^{AS}$ as follows ($\mu_{E^A}$ and $\mu_{E^S}$ computed from embedding table):
\vspace*{-0.8\baselineskip}
\begin{align}
    R^{AS} &= \frac{1}{B-1} (\tilde{E}^A- \mu_{E^A})(\tilde{E}^S- \mu_{E^S})^T\\
    L_{xcorr} &= \frac{1}{D_aD_s} \sum_{i, j} {(R^{AS}_{i,j})^2}
\end{align}
\vspace*{-1.0\baselineskip}

\vspace*{-1.2\baselineskip}
\subsection{Accent conditioned speech synthesis}
\vspace*{-0.5\baselineskip}
We introduce an extra conditioning variable for accent $A$ to RADTTS \cite{radtts} to allow for accent-controllable speech synthesis. We call this model RADTTS-ML, a multilingual version of RADTTS. The following equation describes the model:

\vspace*{-2.0\baselineskip}
\begin{multline}
P_{radtts}(X, \Lambda) = P_{mel}(X | \Phi, \Lambda, A, S) P_{dur}(\Lambda | \Phi, A, S)
\end{multline}

\vspace*{-0.5\baselineskip}
We refer to our conditioning as accent instead of language, because we consider language to be \emph{implicit in the phoneme sequence}. The information captured by the accent embedding should explain the fine-grained differences between how phonemes are pronounced in different languages.

\vspace*{-1.0\baselineskip}
\subsection{Fine-grained frame-level control of speech attributes}
\vspace*{-0.5\baselineskip}
\label{sec:fine_control}
Fine-grained control of speech attributes like $F_0$ and energy $\mathcal{E}$ can provide high-quality controllable speech synthesis\cite{radpp}. We believe conditioning on such attributes can help improve accent and language transfer. During training, we condition our mel decoder on ground truth frame-level $F_0$ and energy. Following \cite{radpp}, we train deterministic attribute predictors to predict phoneme durations $\Lambda$, $F_0$, and energy $\mathcal{E}$ conditioned on speaker $S$, encoded text $\Phi$, and accent $A$. We standardize $F_0$ using the speaker's $F_0$ mean and standard deviation to remove speaker-dependent information. This allows us to predict speech attributes for any speaker, accent, and language and control mel synthesis with such features. We refer to this model as RADMMM, which is described as:
\vspace*{-1.5\baselineskip}

\begin{multline}
P_{radmmm}(X, \Lambda) = P_{mel}(X | \Phi, \Lambda, A, S, F_0, \mathcal{E})\\
P_{F_0}(F_0 | \Phi, A, S) P_{\mathcal{E}}(\mathcal{E} | \Phi, A, S) P_{dur}(\Lambda | \Phi, A, S)
\end{multline}
\vspace*{-1.2cm}
\section{Experiments}\label{sec:experiments}
We conduct our experiments on an open source dataset\footnote{Dataset source, metadata and filelists will be released with source code.} with a sampling rate of 16kHz. It contains 7 different languages (American English, Spanish, German, French, Hindi, Brazilian Portuguese, and South American Spanish). This dataset emulates low-resource scenarios with only 1 speaker per accent with strong correlation between speaker, accent, and language. We use HiFiGAN vocoders trained individually on selected speakers in the evaluation set. 
We focus on the task of transferring the voice of 7 speakers in the dataset to the 6 \emph{other} language and accent settings in the dataset. Herein we refer to RADTTS-ML as RT and RADMMM as RM for brevity.
\vspace*{-0.8\baselineskip}
\subsection{Ablation of Disentanglement Strategies}
\label{sec:ablation}
\vspace*{-0.5\baselineskip}


\begin{figure*}
 \centering
 \begin{subfigure}{0.47\textwidth}
    \centering
    \includegraphics[scale=0.24]{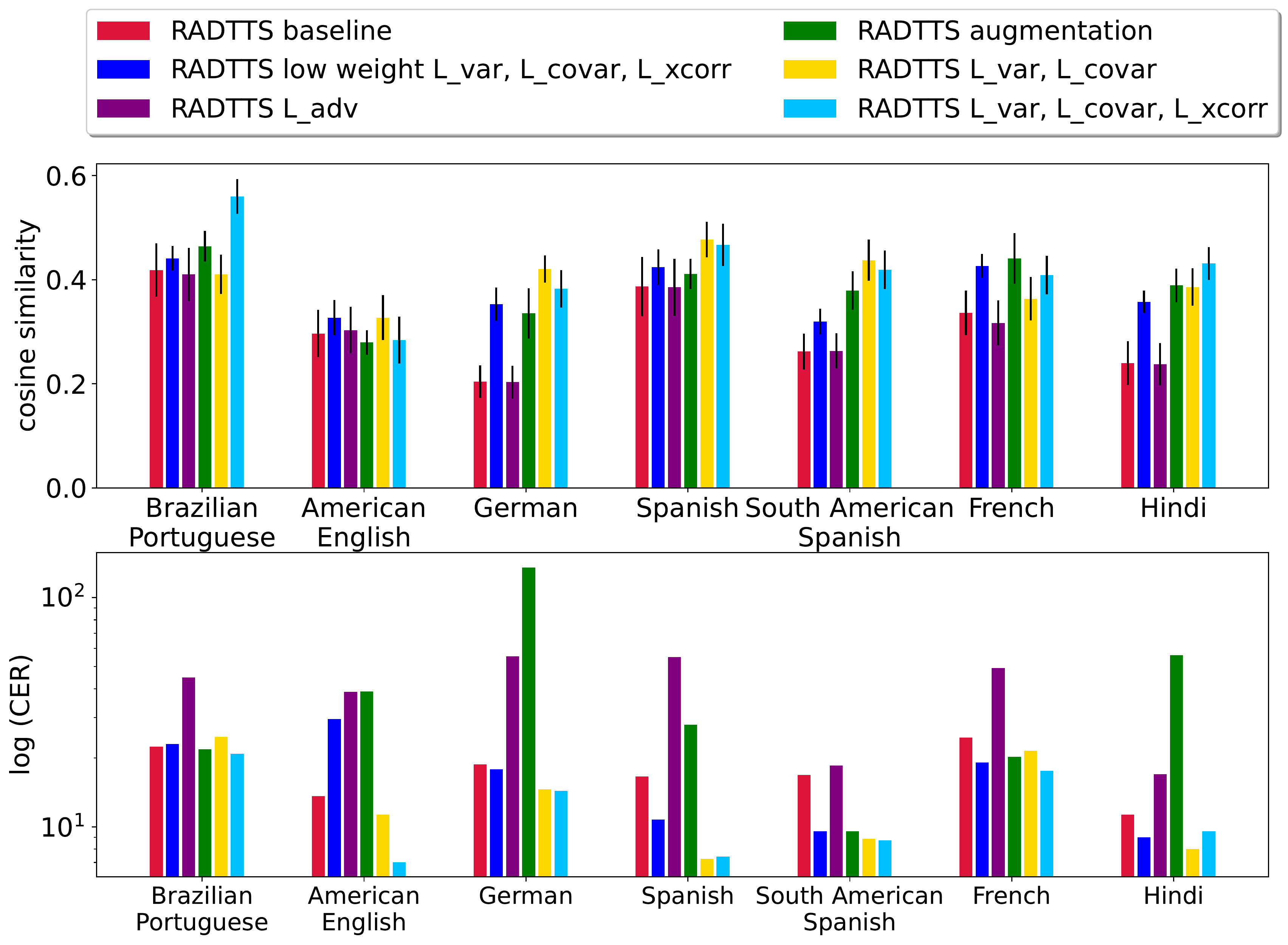}
    \vspace*{-1.5\baselineskip}
    \caption{RADTTS-ML (RT)}
    \label{fig:radttsablation}
\end{subfigure}
\begin{subfigure}{0.495\textwidth}
\centering
  \includegraphics[scale=0.24]{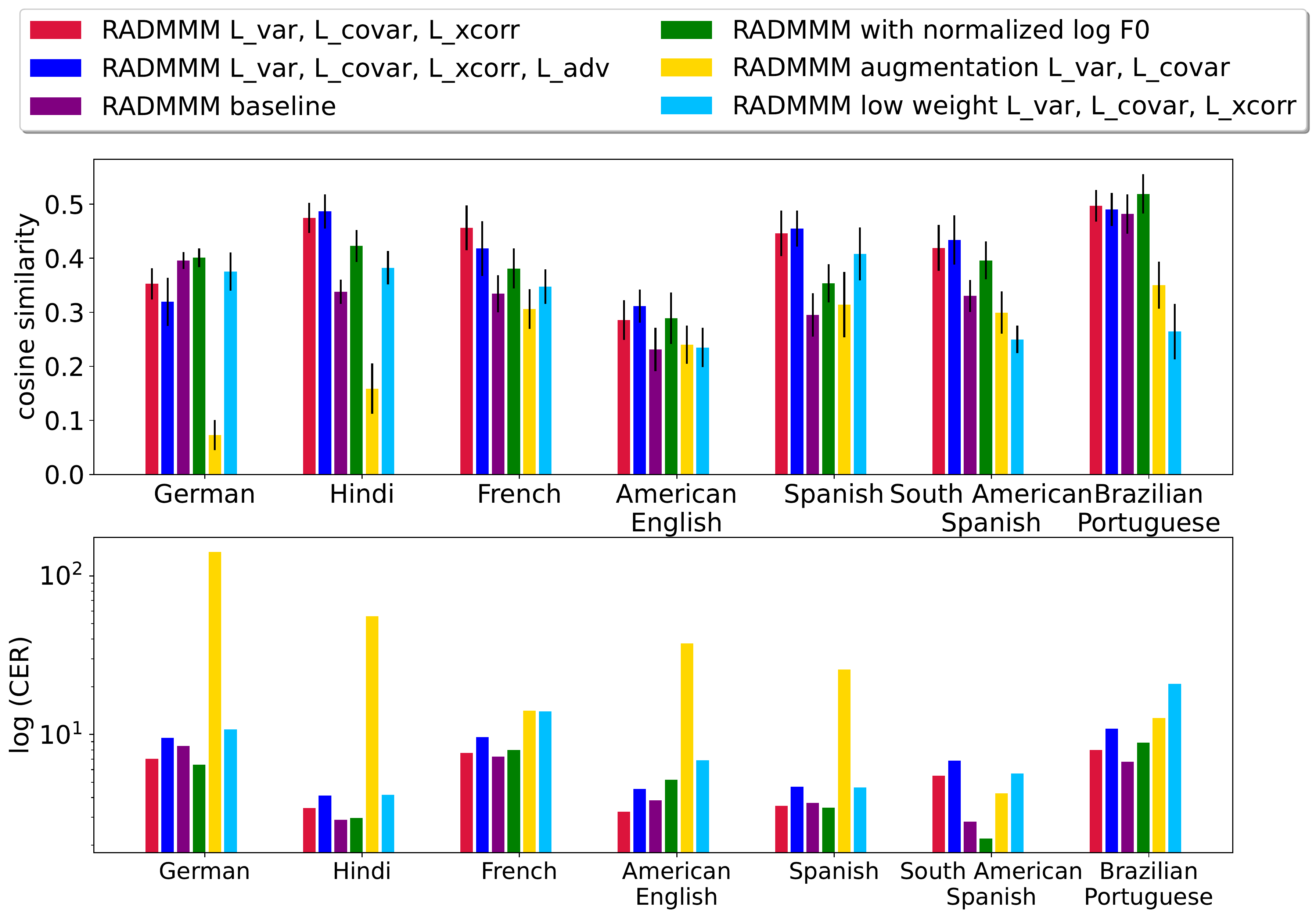}
  \vspace*{-1.5\baselineskip}
  \caption{RADMMM (RM)}
  \label{fig:radmmmablation}
\end{subfigure}
\vspace*{-1.0\baselineskip}
\caption{Comparing speaker cosine similarity and CER of considered disentanglement strategies for every accent.}
\label{fig:internalablationaccentwise}
\vspace*{-1.5\baselineskip}
\end{figure*}

We evaluate the effects of disentanglement strategies on the transfer task by measuring speaker timbre retention using the cosine similarity (Cosine Sim) of synthesized samples to source speaker's reference speaker embeddings obtained from the speaker recognition model Titanet\cite{titanet}. We measure character error rate (CER) with transcripts obtained from Conformer\cite{conformer} models trained for each language. Table \ref{tab:internalablation} and Figure \ref{fig:internalablationaccentwise} demonstrate overall and accent grouped effects of various disentanglement strategies. The \RADTTSML baseline uses the shared text token set, accent-conditioned alignment learning, and no additional constraints to disentangle speaker, text, and accent. The \RADMMM baseline uses this setup with $F_0$ and energy conditioning.
\begin{table}[!ht]
\vspace*{-0.8\baselineskip}
\caption{Ablation results comparing disentanglement strategies using Cosine Sim and CER defined in \ref{sec:ablation}}
\label{tab:internalablation}
\vspace*{-0.8\baselineskip}
\adjustbox{max width=\columnwidth}{%
\begin{tabular}{lcccc}
        \toprule
        \textbf{} & \multicolumn{2}{c}{\textbf{RADTTS-ML}} & \multicolumn{2}{c}{\textbf{RADMMM}} \\
        \textbf{Disentanglement Strategy} & \textbf{Cosine Sim} & \textbf{CER} & \textbf{Cosine Sim} & \textbf{CER} \\
        \midrule
        Baseline (B) & $0.3062 \pm 0.0176$ & 17.7 & $0.3438 \pm 0.0138$ & 5.1 \\
        (B) + normalized $F_0$ pred & N/A & N/A & $0.3946 \pm 0.0143$ & 5.3 \\
        (B) + $L_{adv}$ & $0.3027 \pm 0.0174$ & 39.9 & N/A & N/A \\
        (B) + augmentation & $0.3855 \pm 0.0145$ & 44.3 & $ 0.2174 \pm  0.0131$ & 41.7 \\
        (B) + $L_{var}$ and $L_{covar}$ &  $0.4029 \pm 0.0144$ & 13.7 &  N/A & N/A \\
        (B) + low weight on $L_{var}$, $L_{covar}$ and $L_{xcorr}$ &  $0.3784 \pm 0.0112$ & 17.0 & $0.3232 \pm 0.0154$ & 9.6 \\
        (B) + $L_{var}$, $L_{covar}$ and $L_{xcorr}$ & $0.4217 \pm 0.0156$ & 12.2 & $0.4188 \pm 0.0148 $ & 5.5 \\
        (B) + $L_{var}$, $L_{covar}$, $L_{xcorr}$ and $L_{adv}$ & N/A & N/A & $0.4163 \pm 0.0157$ & 7.2 \\
        \midrule
        \end{tabular}
}

\vspace{-0.8\baselineskip}
\end{table}
\noindent{\textbf{Speaker Adversarial Loss ($L_{adv}$)}}
We observe that the addition of $L_{adv}$ loss to \RADTTSML and \RADMMM does not affect speaker retention when synthesizing the speaker for a target language. However, we observe a drop in character error rate. We believe the gradients from the speaker classifier tend to remove speaker and accent information from encoded text $\Phi$, which affects the encoded text representation leading to worse pronunciation.

\noindent{\textbf{Data Augmentation}}
We use Pratt\cite{Praat} to apply six augmentations: formant scaling down ($\times [0.875 -1.0]$) and up ($\times [1.0 - 1.25]$), scaling $F_0$ down ($\times [0.9 - 1.0]$) and up ($\times [1.0 - 1.1]$), and scaling durations to make samples faster($\times [0.9-1.0]$)) or slower($\times [1.0 -1.1]$). We augment the dataset with transformed audio defining a new speaker identifier, but retaining the original accent. In \RADTTSML, this leads to a significant boost in speaker retention. We believe that creating more speakers per accent enhances disentanglement of accent and speaker. However, in \RADMMM, where $F_0$ is predicted and the model is explicitly conditioned on augmented $F_0$, we observe a significant drop in both speaker retention as well as CER with augmentations, likely due to conditioning on noisy augmented features. 

\noindent{\textbf{Embedding Regularization}}
We conduct three ablations with regularization: one that adds variance ($L_{var}$) and covariance ($L_{covar}$) constraints to the baseline, and two more involving all three constraints ($L_{var}$, $L_{covar}$, $L_{xcorr}$) with small (0.1) and large weights (10.0). We observe an improvement in speaker similarity with the best speaker retention with all three constraints in both \RADTTSML and \RADMMM. Moreover, we observe similar CER to the baselines suggesting similar pronunciation quality. 

Our final models include regularization constraints, but we don't use augmentation and $L_{adv}$ due to worse pronunciation quality and limited success on speaker timbre retention. 
\vspace*{-0.8\baselineskip}
\subsection{Comparing proposed models with existing methods}
\vspace*{-0.5\baselineskip}
We compare our final \RADTTSML and \RADMMM with the Tacotron 2-based model described in \cite{learningfluently}, call it \TC, on the transfer task. We reproduced the model to the best of our ability, noting that training on our data was unstable, possibly due to data quality, and that results may not be representative of the original implementation. We tune denoising parameters\cite{prenger2019waveglow} to reduce audio artifacts from \TC over-smooth generated mels\cite{Flowtron,ren2022revisiting}. We attempted to implement YourTTS\cite{yourtts} but ran into issues reproducing the results on our dataset and hence we don't make a direct comparison to it.

\noindent{\textbf{Speaker timbre retention}}
Table \ref{tab:modelcomparison} shows the speaker cosine similarity of our proposed models and \TC. We observe that both \RADTTSML and \RADMMM perform similarly in terms of speaker retention and achieve better speaker timbre retention than \TC. However, our subjective human evaluation below shows that \RADMMM samples are overall better than \RADTTSML in both timbre preservation and pronunciation.

\begin{table}[ht]
\vspace*{-0.8\baselineskip}
    \small
    \caption{Speaker timbre retention using Cosine Sim (Sec \ref{sec:ablation})}
    \vspace{-0.8\baselineskip}
    \setlength\tabcolsep{4.0pt}
    \centering
    \begin{tabular}{lcc}
        \toprule
        \textbf{Model} & \textbf{Cosine Similarity} \\
        \midrule
        RADTTS-ML (RT) & $0.4186 \pm 0.0154$\\
        RADMMM (RM) & $0.4197 \pm 0.0149$\\
        Tacotron2 (T2) & $ 0.145 \pm 0.0119 $\\
        \midrule
        \end{tabular}
    \label{tab:dataregimes}
    
    \vspace{-1.0\baselineskip}
\label{tab:modelcomparison}
\vspace{-1.0\baselineskip}
\end{table}
\vspace*{-0.5\baselineskip}
\subsection{Subjective human evaluation}
\vspace*{-0.5\baselineskip}
We conducted an internal study with native speakers to evaluate accent quality and speaker timbre retention. Raters were pre-screened with a hearing test based on sinusoid counting. Since MOS is not suited for finer differences, we use comparative mean opinion scores (CMOS) with a 5 point scale (-2 to 2) as the evaluation metric. 
Given a reference sample and pairs of synthesized samples from different models, the raters use the 5 point scale to indicate which sample, if any, they believe is more similar, in terms of accent or speaker timbre, to the target language pronunciation or speaker timbre in reference audio.

\begin{figure}
 \begin{subfigure}{0.49\textwidth}
    \centering
    \includegraphics[width=\linewidth]{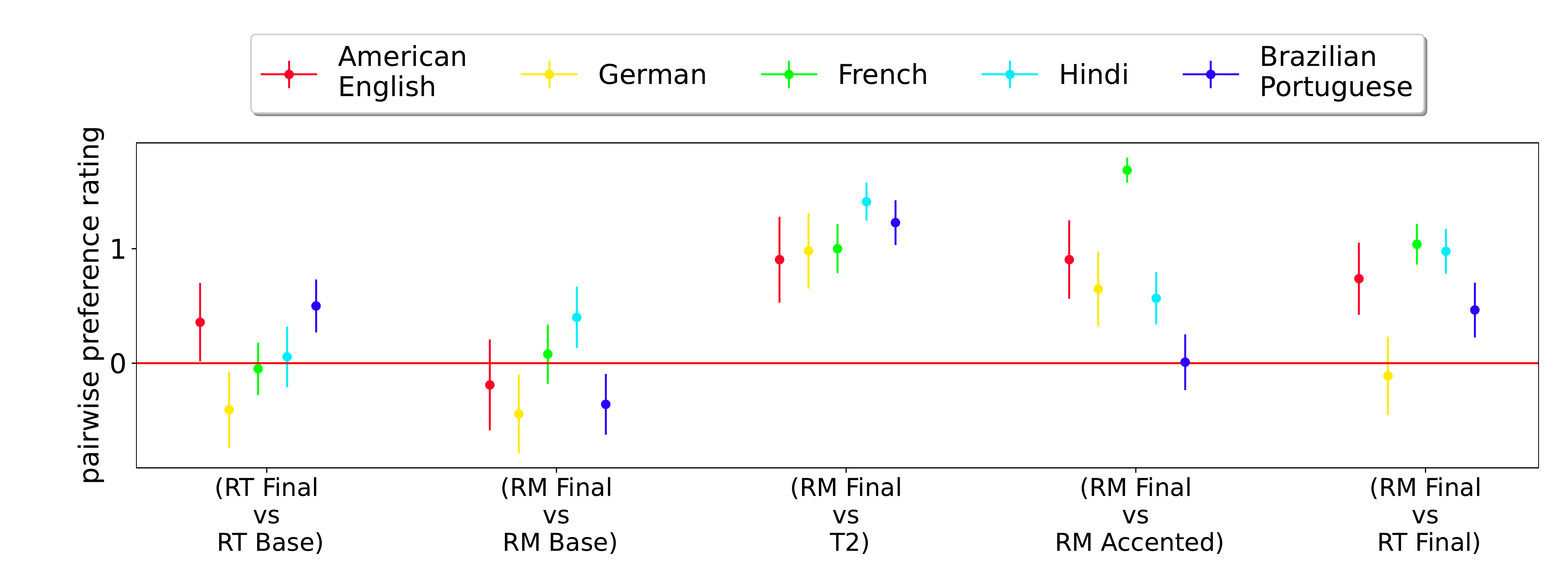}
    \vspace*{-2.0\baselineskip}
\end{subfigure}
\caption{CMOS per accent for model pairs under consideration.}
\label{fig:accentwisecmos}
\vspace*{-1.5\baselineskip}
\end{figure}

\noindent{\textbf{Accent evaluation:}}
We conduct accent evaluation with native speakers of every language. 
Fig \ref{fig:accentwisecmos} shows the preference scores of native speakers with $95\%$ confidence intervals in each language for model pairs under consideration. Positive mean scores imply that the top model was preferred over the bottom model within the pair. Given limited access to native speakers, we show results for only 5 languages. We observe that there is no strong preference between \RADTTSML final and its baseline in terms of accent quality. We find similar results for \RADMMM final and its baseline, suggesting that accent and pronunciation are not compromised by our suggested disentanglement strategies. To evaluate controllable accent, we synthesize samples from our best model (\RADMMM final) for every speaker in languages other than the speaker's native language. Samples using non-native language and accent are referred to as \RADMMM final, and samples with the new language but native accent are referred to as \RADMMM accented. Raters preferred samples using the target accent (RM final) over the source speaker's accent (RM accented), indicating the effectiveness of accent transfer. Finally, \RADMMM final is preferred over \TC in terms of accent pronunciation.


\noindent{\textbf{Speaker timbre evaluation:}} 
Table \ref{tab:cmosresults} shows CMOS scores with $95\%$ confidence intervals. First, we observe that in both \RADTTSML and \RADMMM, the final models with disentanglement strategies applied are preferred over baseline models in terms of speaker timbre retention. \RADMMM accented synthesis (RM accented) is rated as having similar speaker timbre as native accent synthesis with \RADMMM (RM final), indicating that changing accent doesn't change speaker timbre in \RADMMM, thus showcasing the disentangled nature of accent and speaker. Finally, RM final is preferred over \TC in terms of speaker timbre retention on transferring speaker's voice to target language.

\noindent{\textbf{Effects of control with $F_0$ and $\mathcal{E}$:}} Comparing \RADMMM final with \RADTTSML final, we see that \RADMMM is preferred for most languages except German, indicating that explicit conditioning on $F_0$ and energy results in better pronunciation and accent. Moreover, as illustrated in Table \ref{tab:internalablation}, \RADMMM final achieves a better CER than \RADTTSML final. Table \ref{tab:cmosresults} demonstrates that explicit conditioning on $F_0$ and energy in \RADMMM results in much better speaker timbre retention compared to \RADTTSML. \RADMMM results in the best speaker retention, accent quality and pronunciation among our models.

\vspace*{-0.8\baselineskip}
\begin{table}[ht]
    \small
    \caption{CMOS for speaker timbre similarity.}
    \vspace{-1.0\baselineskip}
    \setlength\tabcolsep{1.5pt}
    \centering
    \begin{tabular}{lc}
        \toprule
        \textbf{Model Pair} & \textbf{CMOS} \\
        \midrule
        RT Final vs RT Base & $0.300 \pm 0.200 $ \\
        RM Final vs RM Base & $0.750 \pm 0.189 $ \\
        RM Final vs RT Final & $0.733 \pm 0.184$ \\
        RM Final vs RM Accented & $0.025 \pm 0.199$ \\
        RM Final vs T2 & $1.283 \pm 0.144$ \\
        \midrule
        \end{tabular}

\label{tab:cmosresults}
\vspace{-1.0\baselineskip}
\end{table}

\vspace{-1.0\baselineskip}
\section{Conclusion}\label{sec:discussion}
\vspace*{-1.0\baselineskip}
We present a multilingual, multiaccented and multispeaker TTS model based on RADTTS with novel modifications. We propose and explore several disentanglement strategies resulting in a model that improves speaker, accent and text disentanglement, allowing for synthesis of a speaker with closer to native fluency in a desired language without multilingual speakers. Internal ablation studies indicate that explicitly conditioning on fine-grained features ($F_0$ and $\mathcal{E}$) results in better speaker retention and pronunciation according to human evaluators. Our model provides an ability to predict such fine-grained features for any desired combination of speaker, accent and language and
user studies show that under limited data constraints, it improves pronunciation in novel languages. Scaling the model to large-resource conditions with more speakers per accent remains the subject of future work. 

\vfill\pagebreak

\bibliographystyle{IEEEbib}
\bibliography{main}

\newpage
\end{document}